\input{aipcheck}

\documentclass[
    ,final            
  ]
  {aipproc}

\layoutstyle{6x9}

\newcommand{\be}{\begin{eqnarray}}
\newcommand{\ee}{\end{eqnarray}}

\def\lsim{\mathrel{\rlap{\lower4pt\hbox{\hskip1pt$\sim$}}
\raise1pt\hbox{$<$}}}               
\def\gsim{\mathrel{\rlap{\lower4pt\hbox{\hskip1pt$\sim$}}
\raise1pt\hbox{$>$}}}               

\begin{document}

\title{Light Cone Dynamics and EMC Effects in the Extraction of
  $F_{2n}$ at Large Bjorken $x$.}

\classification{11.80.-m,13.60.-r,13.85.Ni}
\keywords      {deep-inelastic scattering, light-cone dynamics, neutron structure function}

\author{Misak M. Sargsian}{
  address={Department of Physics, Florida International University, Miami, FL 33199}
}
 
\begin{abstract}
We discuss  theoretical issues related to the extraction of deep inelastic~(DIS) structure function of neutron 
from inclusive DIS scattering off the deuteron at large Bjorken $x$.  Theoretical justification is given to 
the consideration of  only $pn$ component of the deuteron wave function and   
consistency  with both the baryonic number  and light-cone momentum conservation sum rules.
Next we discuss the EMC type effects and argue that  in all cases relevant to the nuclear DIS reactions at large $x$ 
the main issue is the medium modification of the properties of bound nucleon rather than the non-nucleonic 
components like pions.   We give brief description of the color screening model of EMC and within this model 
we estimate uncertainties in the extraction of the neutron DIS structure function at large $x$.  We emphasize 
also that these uncertainties are rather "model independent" since any theoretical framework accounting for 
the medium modification is  proportional to the magnitude of the virtuality of bound nucleon which increases with 
an increase of $x$.
 \end{abstract}

\maketitle
 
\section{Introduction}

Since the pioneering  experiments  on deep inelastic 
electron scattering off the deuteron at SLAC in late 70's\cite{Bodek} the investigation of the 
partonic distribution functions~(PDFs) of the neutron at large x is one of the most important 
topics in the high x QCD studies.

The main target of the choice for the  extraction of the neutron PDFs is the deuteron and 
as the studies progressed it  become more and more clear that  several issues purely in  
nuclear physics nature  should be addressed for   successful exploration of DIS structure of the 
neutron.

Historically, due to its simplicity the main reaction considered was the inclusive $d(e,e')X$ scattering 
of the deuteron. However new generation of semi-inclusive experiments\cite{deeps,bonus6,bonus12}  present 
completely new framework in studies of neutron PDFs\cite{mss97,polext,ciofi1,wim1,wim2}.

In inclusive DIS scattering several issues pertaining to the deuteron and the nature of the inclusive
 scattering are important for  unambiguous interpretation of the neutron DIS data.
These includes  appropriate description of the scattering process off the relativistic 
 bound system, effects due to modification of bound nucleon structure as well final state interaction.

In this report we focus on two issues such as light-cone description of the inclusive DIS process 
involving  deuteron and medium modification of neutron PDFs.

\section{Light Cone Description of the Inclusive DIS Scattering}

 Since partons  have meaningful interpretation only in the infinite momentum frame or in the light cone\cite{Feynman} the 
 selfconsistent  interpretation  of DIS scattering off the nuclei in therms of partonic degrees of freedom requires  the 
 description of the nuclei in the infinite momentum or light cone reference frame. 
  In this case a bound nucleon is  described by light-cone momentum $\alpha$ which is  Lorentz  invariant  quantity boosted in the 
  direction of  infinite momentum and has a meaning of the momentum fraction of the nucleus carried by the bound nucleon.
  
  In the case for  the deuteron assuming that it consists of only  proton and neutron we can express the relative momentum of $pn$ system 
  in the light-cone through  the momentum fraction $\alpha$ as follows\cite{FS81}:
  \begin{equation}
  k = \sqrt{{m_N^2 + p_t^2\over \alpha(2-\alpha)} - m_N^2}
\label{k}
  \end{equation}
  where  momentum fractions are normalized in such a way that for stationary nucleon $\alpha=1$.  The above defined momentum $k$ allows us 
  to estimate the limit at which  one can consider  the deuteron as consisting of proton and neutron only, 
  For this in Fig.1 we present the Bjorken - $x$ dependence of $k$ for typical DIS scattering kinematics at large $x$ and compare it with 
  the three momentum of the nucleon as it enters in the lab frame description of the deuteron.   From the figure we observe that 
  at large $x$ the relative momentum of $pn$ system in the light cone is consistently less than the the one defined in the Lab frame.
  This situation is important from the point of view of justification of the approximation in which deuteron consists of only   proton and neutron. 
  Based on the recent observation\cite{srcrev}  that   the nucleonic component  in the isosinglet $pn$ short range correlation  dominates 
  in the wave function  till    $\sim 650$~MeV/c relative momenta we were able to estimate that for $Q^2\ge 5$~GeV$^2$ and practically for
  whole range of large Bjorken $x \le 1$ the non-nucleonic components can be safely neglected in the ground state wave function of the 
  deuteron.  
  
\begin{figure}
 \includegraphics[height=.3\textheight]{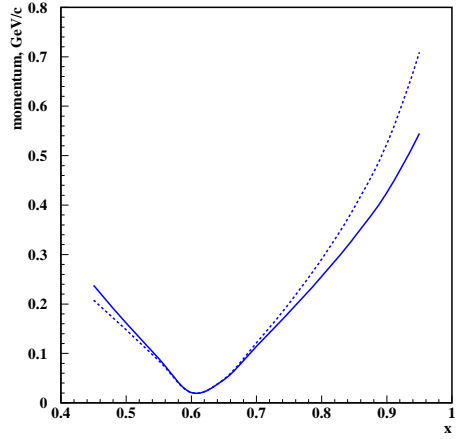}
  \caption{The dependence of minimal  light-cone  $k$~(solid line)  and lab frame $p$ ~(dashed line)  relative momenta of 
the   $pn$ system  in the deuteron on the Bjorken $x$.  Calculations  are done for $Q^2=5$~GeV$^2$ and $w_N = 2$~GeV.} 
\end{figure}
  Constraining ourselves  only by  two-nucleon component of the deuteron wave function within LC approximation for the DIS structure 
  function of the deuteron we obtain\cite{FS88,SSS02}
  \be
      F_2^A(x, Q^2) & = & \sum_{N = 1}^2 \int {d\alpha \over \alpha^2} d^2p_{\perp} ~ 
      \rho^{LC}_N(\alpha, p_{\perp}) ~ F_2^N(\tilde{x}, Q^2) {\nu \over 
      \tilde{\nu}} \nonumber \\ 
      & \times & \left[ ({M_d \over 2m_N})^2 (1 + \mbox{cos} \delta)^2 (z + 
      \alpha_q {m_N \nu' \over Q^2})^2 + {p_{\perp}^2 \over 2 M_N^2} 
      \mbox{sin}^2\delta \right], 
      \label{F2LC} 
 \ee
where  
 \be
      \tilde{x} & = & {Q^2 \over 2m_N \tilde{\nu}},  \ \ \ \   \tilde{\nu}  =  {w_N^2 + Q^2 - m_N^2 \over 2m_N},  \nonumber \\
      w_N^2 & = & Q^2 + {1 \over 2}{M_d \over 2}(p_+ \alpha_q + z q_+) + {M_d 
      \over 2} p_+z - p_{\perp}^2, \nonumber \\
      \nu' & = &  {1 \over 2m_N} (p_+ q_- + p_- q_+) = {M_d \over 2m_N} \left[
      p_+ \alpha_q + q_+ z \right],
      \label{pplus}
 \ee
and $p_\pm  = E \pm p_z$  where $z$ axis is defined in the direction of $\vec q$.
The  light-cone density matrix   $\rho^{LC}_N(\alpha, p_{\perp})$  can be expressed through the deuteron wave function as follows\cite{FS81}:
\begin{equation}
\rho^{LC}_N(\alpha, p_{\perp})= {E_k |\Psi_d(k)|^2 \over 2 - \alpha}
\label{lc_denfun}
\end{equation}
where $E_k = \sqrt{m_N^2+k^2}$   and the momentum $k$ is defined according to Eq.(\ref{k}).
The light-cone density matrix defined above satisfies two sum rules:  From baryon charge conservation one has
 \be
       \int {d\alpha \over \alpha} d^2p_{\perp} ~ \rho^{LC}_N(\alpha, p_{\perp}) = 1, 
       \label{bchc}
 \ee
while the momentum sum rule requires
 \be
       \int {d\alpha \over \alpha} d^2p_{\perp} ~ \alpha \rho^{LC}_N(\alpha,  p_{\perp})  = 1.
       \label{msm}
 \ee
 The above two relations are necessary conditions for self-consistency  if one excludes any non-nucleonic component in 
 the deuteron wave function.  In this respect it is interesting that within approximations  in which the struck nucleon 
 is treated as virtual in the lab frame of the scattering process~(generally referred as virtual nucleon~(VN) approximation) 
 (see e.g.\cite{FGMSS95,FSS97,treview,MS09,wim1,wim2}) the momentum sum rule of Eq.(\ref{msm}) is not satisfied and:
 \be
       \int {d\alpha \over \alpha} d^2p_{\perp} ~ \alpha \rho^{VN}_N(\alpha,  p_{\perp})  < 1.
       \label{msmvn}
 \ee
Such a result can be interpreted as  missing momentum fraction  being distributed to the unaccounted degrees of freedom such 
as pions.
Note that the last sum rule is not directly satisfied in the $VN$ model, but it can be restored if mesonic degrees of freedom  are 
introduced explicitly (see e.g. Ref.\cite{KMK}.

The account of both sum rules, given above,  within light-cone approximation  leads to a prediction for the $F_2^A / F_2^N$ ratio which qualitatively 
contradicts the $EMC$ effect for $x \gsim 0.5$ (Fig.2).  This situation however indicates that the next step in the description of 
DIS scattering off  the nucleus should be the account of nuclear medium modifications of  the structure functions of bound nucleon.

\section{Medium Modification Effects}

The discovery of the  $EMC$ effect at large $x$ has triggered a huge theoretical effort which has led to the development of a large number of models 
(see, e.g., Refs. \cite{Close,Jaffe,FS85,Pirner,FJM,SmithMiller}).   One can divide  these models in two groups: one in which  the effect 
is  due to  the missed non-nucleonic  component  (such as pion degrees of freedom) and the other  group 
in which  EMC effects are due to modification of  the properties of bound nucleons.  According to our discussion above we believe that the first group
contributes little in the DIS kinematics at large $x$. Moreover it can be shown that in the LC approximation even if pions 
will carry sizable light cone momentum fraction it will not be accessible in  DIS measurement\cite{Miller:1997} .

The important characteristics of  the models in the second group is that the extent  of modification  depends on   the magnitude of    
virtuality of the  bound nucleon.    In this case as it follows from Fig.1 one expects that with an increase of $x$  the medium modification 
of the DIS structure function will be more and more important.

To estimate the expected magnitude of  the effects in  the extraction of  large $x$  neutron  structure function from inclusive DIS scattering off the 
deuteron we calculate the medium modification based on one of "second group" models:  the color screening model of Ref.\cite{FS85,FS88}.
 
 This model is based on the observation that  the most significant EMC effect  is observed in the range of  $x$  corresponding to 
 high momentum components of the quark distribution in the nucleon and therefore the $EMC$ effect is expected to be mostly sensitive to 
 the nucleon  wave function configurations where three quarks are likely to be close together. 
 Such  small size configurations are referred as  point-like configurations ($PLC$). 
 It is then assumed that for large $x$ the dominant contribution to $F_2^N(x,Q^2)$ is given by $PLC$ of partons which, due to color screening, 
 interact weakly  with the other nucleons.  Because  of this the optimally bound configuration of nucleons will have suppressed contribution from the 
 PLC component of nucleon wave function.
 The suppression of $PLC$ in a bound nucleon is assumed to be the main source of the $EMC$ effect in inclusive $DIS$  and 
 the suppression factor is calculated in perturbation series of the parameter:
 \be
       \kappa = \left| {\langle U_A \rangle \over \Delta E_A}  \right| 
       \label{plc1}
 \ee
where $\langle U_A \rangle$ is the average potential energy per nucleon  and $\Delta E_A \approx M^* - M \sim 0.6 \div 1 ~ GeV$ is the typical energy for nucleon excitations within the nucleus.

\begin{figure}
 \includegraphics[height=.35\textheight,width=0.6\textwidth]{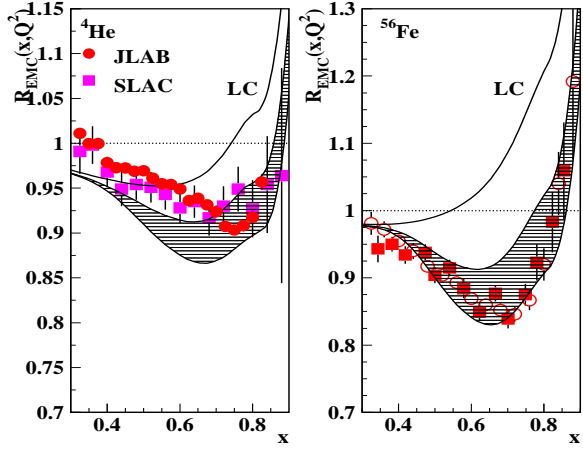}
  \caption{The $x$ dependence of EMC ratios. Solid line light-cone approximation without medium modifications.
  Dashed area corresponds to LC calculations with EMC effects calculated according to color screening model. 
  Largest  effect corresponds to $\Delta E_A = 0.6$~GeV and smallest - $\Delta E = 1$~GeV.  The data are from 
  Refs.\cite{Gomez,JLab}.} 
\end{figure}

To calculate  the deformation of the quark wave function in the bound nucleon due to suppression of the probability of $PLC$ in the bound 
nucleon and then to account for it in the calculation of $F_2^A(x, Q^2)$  one needs to introduce the nuclear potential with explicit  
quark degrees of freedom in it: $V(R_{ij}, y_i, y_j)$. Then using this potential to reevaluate the potential energy $U$ that enters 
into Schr\"odinger equation for  the nuclear ground state  wave function in the form:
   \be
       U(R_{ij}) = \sum_{y_i, y_j, \widetilde{y}_i, \widetilde{y}_j} 
       \langle \phi_N(y_i) \phi_N(y_j) \mid V(R_{ij}, y_i, y_j, 
       \widetilde{y}_i, \widetilde{y}_j) \mid \phi_N(\widetilde{ y}_i) 
       \phi_N(\widetilde{ y}_j) \rangle, 
       \label{plc3}
   \ee
where $\phi_N(y_i)$ is the free nucleon wave function. Using for the unperturbed nuclear wave function the solution of the Schr\"odinger equation with $U(R_{ij})$, one can treat $(U - V) / \Delta  E_A)$, as a small parameter to calculate the dependence of the probability to find a nucleon in a $PLC$ on the momentum of the nucleon inside the nucleus.  Such a calculation allows to estimate the suppression of the probability to find a $PLC$ in a bound nucleon as compared to the similar probability for a free nucleon. In the $DIS$ cross section the $PLC$ suppression can be represented as a suppression factor $\delta_A(k^2)$ which is multiplicative to the nucleon structure function $F_2^N(\tilde{x}, Q^2)$ in the $LC$ convolution formula 
of  Eq.(\ref{F2LC})\cite{FS85}
 \be
        \delta_A(p^2) = {1\over (1+\kappa)^2} = {1 \over [1 + (p^2 / M + 
        2\epsilon_A) / \Delta E_A]^2}, \nonumber \\ 
        \label{eq:gamma}
 \ee
where $k$ is the momentum of the bound nucleon in the light cone.
Finally the $x$ dependence of the suppression effect is based on the assumption that the $PLC$ contribution in the nucleon wave function is negligible at $x \lsim 0.3$, and gives the dominant contribution at $x \gsim 0.5$. We use a simple linear fit to describe the $x$ dependence between these two values of $x$ \cite{FSS90}.    

Using the above estimate of the suppression factor  we present in Fig.2  the comparison of our calculations for  
$\Delta E_A \approx M^* - M \sim 0.6 \div 1 ~ GeV$  with the old SLAC\cite{Gomez} and new JLab\cite{JLab} data.
These comparisons also demonstrate that the LC calculation without  medium modification disagrees strongly  
with the measured EMC ratios. 

\begin{figure}
 \includegraphics[height=.3\textheight]{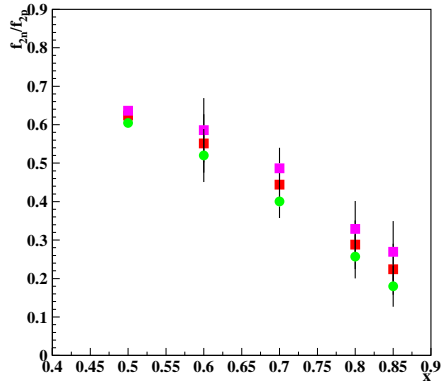}
  \caption{The $x$ dependence of  extracted ${F_{2n}\over F_{2p}}$ ratios from the DIS deuteron data\cite{DBS}.
  Circles - LC model, lower squares minimal EMC effect upper squares maximal EMC effects. }
\end{figure}

\section{Extraction of the Neutron DIS Structure Function}

Within the above described theoretical model  we perform the extraction of neutron DIS stricture function $F_{2n}(x)$ with the similar procedure
used  in Ref.\cite{Bodek}.   In the estimates in which EMC effects are taken into account we first modify the proton structure  function,  smear it 
by Fermi motion and then 
subtract  from the deuteron data. After correcting by  nucleon motion effects we modify back the extracted neutron structure functions 
to reconstruct "free"   $F_{2n}$ for neutron.
The results are presented in Fig.3, which indicates that larger is $x$ more important are nuclear modification effects due to 
large virtuality of nucleons (see Fig.1)  involved in DIS scattering.

 \section{Conclusion and Outlook}

It can be shown that other models of EMC based on the modification of structure function of nucleon exhibit  similar 
dependence  on the momentum (virtuality) of bound nucleon\cite{SSS02,hnm}. As a result one expects that the uncertainty in 
the extraction of large $x$ DIS structure functions  of the neutron from  inclusive scattering off the deuteron to be rather
"model independent".   This emphasizes further the urgency of  understanding the origin and extent of EMC effects  relevant 
to large $x$ kinematics.

\begin{theacknowledgments}
Many thanks to Mark Strikman and Leonid Frankfurt for many years of collaboration on studies of nuclear DIS and 
EMC effects.  This work is supported  by  the United States  Department of Energy  Grant under Contract DE-FG02-01ER-41172.

\end{theacknowledgments}
 \bibliographystyle{aipproc}  
\IfFileExists{\jobname.bbl}{}
 {\typeout{}
  \typeout{******************************************}
  \typeout{** Please run "bibtex \jobname" to optain}
  \typeout{** the bibliography and then re-run LaTeX}
  \typeout{** twice to fix the references!}
  \typeout{******************************************}
  \typeout{}
 }


\end{document}